\documentclass[sigconf, screen, anonymous=False]{acmart}
\PassOptionsToPackage{dvipsnames}{xcolor}
\PassOptionsToPackage{colorlinks=true,linkcolor=RubineRed,breaklinks=True,citecolor=RubineRed}{xcolor}

\usepackage{enumitem}
\setlist[itemize]{noitemsep, topsep=0pt}

\usepackage[most]{tcolorbox}
\usepackage{adjustbox}
\usepackage[linesnumbered,algo2e,ruled]{algorithm2e}
\usepackage{multirow}
\usepackage{booktabs,subcaption,dcolumn}
\usepackage{tikz}
\usepackage{enumitem}
\usepackage{tabularx}
\usepackage{MnSymbol}

\usepackage{pifont}
\usepackage{svg}
\usepackage{soul}
\usepackage{amsmath}
\usepackage{dirtytalk}
\usepackage{subcaption}
\usepackage{graphicx}
\usepackage{setspace}
\usepackage[font=small]{caption}
\usepackage{booktabs} 
\usepackage[titletoc,toc,title]{appendix}
\usepackage{svg}
\usepackage{amsfonts}
\usepackage{xurl} 

\usepackage{diagbox}
\usepackage{indentfirst}

\usepackage{flushend}
\hypersetup{draft}

\usepackage{CJKutf8}
\usepackage{amsmath}
\usepackage{amsfonts} 

\usepackage{colortbl}

\setcopyright{acmlicensed}
\copyrightyear{2025}
\acmYear{2025}
\acmDOI{10.1145/3696673.3723067}
\acmConference[ACMSE 2025]{2025 ACM Southeast Conference}{April 24--26, 2025}{Cape Girardeau, MO, USA}
\acmBooktitle{2025 ACM Southeast Conference (ACMSE 2025), April 24--26, 2025, Cape Girardeau, MO, USA}
\acmISBN{979-8-4007-1277-7/2025/04}

\AtBeginDocument{%
  \providecommand\BibTeX{{%
    \normalfont B\kern-0.5em{\scshape i\kern-0.25em b}\kern-0.8em\TeX}}}

\newcolumntype{?}{!{\vrule width 1.5pt}}

\newtcolorbox{cooltextbox}[1][]{%
    colback=black!5,
    colframe=black!5,
    notitle,
    sharp corners,
    borderline west={0pt}{0pt}{red!80!black},
    enhanced,
    breakable,
    left=0pt,
    right=0pt,
    top=0pt,
    bottom=0pt
    }

\definecolor{table_1_blue}{RGB}{0, 0, 204} 
\definecolor{table_1_green}{RGB}{199, 21, 143} 
\newcommand\revision[1]{%
  \bgroup
  \hskip0pt\color{blue!80!black}%
  #1%
  \egroup
}

\begin{document}
\pagestyle{empty}

\title{How Do Mobile Applications Enhance Security? An Exploratory Analysis of Use Cases and Provided Information}


\author{Irdin Pekaric}
\orcid{0000-0002-0706-3202}
\affiliation{%
  \institution{University of Liechtenstein}
  \city{Vaduz}
  \country{Liechtenstein}
}
\email{irdin.pekaric@uni.li}

\author{Clemens Sauerwein}
\affiliation{%
  \institution{University of Innsbruck}
  \city{Innsbruck}
  \country{Austria}}
\email{clemens.sauerwein@uibk.ac.at}

\author{Simon Laichner}
\affiliation{%
  \institution{University of Innsbruck}
  \city{Innsbruck}
  \country{Austria}}
\email{simon.laichner@student.uibk.ac.at}

\author{Ruth Breu}
\affiliation{%
  \institution{University of Innsbruck}
  \city{Innsbruck}
  \country{Austria}}
\email{ruth.breu@uibk.ac.at}

\renewcommand{\shortauthors}{Pekaric et al.}

\begin{abstract}

\noindent The ubiquity of mobile applications has increased dramatically in recent years, opening up new opportunities for cyber attackers and heightening security concerns in the mobile ecosystem. As a result, researchers and practitioners have intensified their research into improving the security and privacy of mobile applications. At the same time, more and more mobile applications have appeared on the market that address the aforementioned security issues. However, both academia and industry currently lack a comprehensive overview of these mobile security applications for Android and iOS platforms, including their respective use cases and the security information they provide. 

To address this gap, we systematically collected a total of 410 mobile applications from both the App and Play Store. Then, we identified the 20 most widely utilized mobile security applications on both platforms that were analyzed and classified. Our results show six primary use cases and a wide range of security information provided by these applications, thus supporting the core functionalities for ensuring mobile security.
\end{abstract}


%
%




\settopmatter{printfolios=true}

\maketitle

\section{Introduction}
\label{sec:introduction}
\noindent
In the course of the rapid digitalization of society and the economy, the spread of mobile applications has increased significantly in recent years \cite{hinze2023study}. The number of available mobile applications on the Play Store and the App Store has grown over two million\footnote{\url{https://42matters.com/google-play-statistics-and-trends}, \url{https://42matters.com/ios-apple-app-store-statistics-and-trends}}. As a result, mobile devices and applications have become lucrative targets for cyber attacks. This is reflected in a staggering 50\% increase in attacks on mobile devices between 2022 and 2023, culminating in an incredible 33.8 million attacks worldwide in 2023\footnote{\url{https://www.kaspersky.com/about/press-releases/2024_attacks-on-mobile-devices-significantly-increase-in-2023}}. 

This digital revolution and the ubiquity of smartphones have created a vast attack surface for adversaries. Each device can potentially be used as a gateway to sensitive personal and corporate information. The potential attack vectors can range from phishing schemes that exploit the limited security features of mobile devices, to malware disguised as legitimate applications. Moreover, the implications of this surge in mobile-targeted attacks extend beyond individual users, potentially compromising entire organizational networks. As our reliance on mobile technology continues to grow and evolve, the urgency to develop robust, adaptive security measures that can keep pace with the evolving threat landscape is of utmost importance.

Both research and practice are looking into the security and privacy aspects of mobile applications and platforms. For example, some research focuses on the security and privacy of mobile applications through detailed analyses that consider static and dynamic technologies \cite{li2021mobile,papageorgiou2018security}, while other studies examine the security awareness of users in connection with the use of mobile applications \cite{zeybek2019study,tao2020identifying}. However, due to the large number and diversity of the applications, it is difficult to tell what the available apps are used for. For instance, it can be argued that these applications provide a solid foundation for sharing threat intelligence data due to the fact that they can provide timely information for the users (e.g. attack and vulnerable asset information).

However, academia and industry currently face the challenge of lacking a comprehensive overview of mobile security applications for Android and iOS platforms. This includes detailed insights into their respective use cases and the specific type of security information they provide, which was done in other domains such as automotive {\cite{DBLP:journals/csi/PekaricSHF21, pekaric2019applying}, self-adaptive system \cite{witte2022adaptive, pekaric2023adaptive} and web \cite{sauerwein2019analysis} domains. Such comprehensive information can serve as a backbone for future developments and provide a foundation for more targeted research in the domain of mobile security applications. Thus, we empower users to make informed decisions about which security tools best suit their needs. Understanding the types of security information these applications provide allows users to better protect themselves against evolving threats. Moreover, this study lays the groundwork for developers and researchers to identify gaps in current mobile security offerings, potentially leading to the creation of more effective and user-centric security solutions. 

To the best of our knowledge, no other work has compared mobile applications' actions across different platforms until now. To address this gap, we investigate the two aforementioned aspects by focusing on the following two research questions: 
\begin{itemize}
    \item [\textbf{RQ1}] What are the use cases of mobile security applications?
    \item [\textbf{RQ2}] What type of security information is provided by mobile security applications?
\end{itemize}

To answer these research questions, we conducted an exploratory systematic analysis of popular mobile security applications in the App and Play Store. We systematically identified a total of 410 applications, of which we analyzed the 20 most utilized applications in both the App Store and the Play Store. These were retrieved using a custom tool we developed and then manually classified according to dimensions described in Sections \ref{subsec:usecase} and \ref{subsec:security_information}. 

According to the results of the study, we identified six different common use cases including  \textit{Security Education \& Training}, \textit{Antivirus Protection}, \textit{Secure Browsing}, \textit{Privacy Management}, \textit{Network \& Device Management}, and \textit{Secure Communication}. In regards to the type of security information the applications provide, \textit{countermeasure}, \textit{asset}, \textit{threat} and \textit{risk information} were found to be the most prevailing type of information.

\noindent
\textsc{\textbf{Contributions.}} We want to emphasize the role of mobile security applications and the security information they provide. After shaping the problem space in Section \ref{sec:related}:
\begin{itemize}[leftmargin=*]
    \item First, this is done through a custom-developed tool, we collect 410 security applications (Section \ref{subsec:search});
    \item Next, we select the top 20 most used applications for each store and rigorously review these (see Sections~\ref{subsec:selection} and \ref{subsec:extraction});
    \item Finally, we analyze their use cases and type of provided security information\footnote{The original motivation behind this study was to investigate what type of security information is shared by mobile cyber threat intelligence applications. However, we were not able to identify these particular types of applications so we focused on general mobile security applications in order to determine which type of security information is present since these can be utilized as a part of the broader cyber threat intelligence processes as demonstrated by Mavroeidis and Bromander \cite{mavroeidis2017cti}. Additionally, this approach can provide insights into the future development of mobile cyber threat intelligence applications.} (Section \ref{sec:results}).
\end{itemize}

The rest of the paper also includes Section~\ref{sec:discussion} that discusses key findings and limitations of the research at hand, while Section~\ref{sec:conclusions} concludes our work and provides an outlook on future work. For the purpose of open science, we release all our resources at: \url{https://github.com/irdin-pekaric/MobileAppACMSE2025}.

\section{Related Work}
\label{sec:related}
\noindent

\noindent The existing body of related work on mobile security applications and conventional security tools is highly limited. This limitation underscores the necessity of exploring the related works that cover both domains in order to gain a comprehensive understanding of the state-of-the-art. However, we consider related works that focus on analyzing security aspects of a single application out of scope and we only consider the ones that target multiple applications. This is due to the fact that this work investigates multiple applications. Accordingly, we identified studies investigating \textit{conventional security tools}, \textit{privacy of mobile applications}, and \textit{security awareness of mobile applications' users}.

\textit{Conventional security tools} - Two notable studies exist in this particular realm. Kuehn et al. \cite{kuehn2022notion} conducted a categorization of security tools, focusing on seven general dimensions. Similarly, the paper by Curphey and Arawo \cite{curphey2006web} classified web application security tools based on high-level types, which provided insights into the broader landscape of security tools. However, compared to our study, the classifications presented in both of these papers were not applied to the domain of mobile applications, leaving a gap in understanding the ecosystem of mobile security tools.   

\textit{Privacy of mobile applications} - The most addressed security aspect of mobile applications in previous works is privacy. Li et al. \cite{li2021mobile} conducted a study on the detection and analysis of personal information security wherein they applied static analysis, dynamic analysis, and manual review to detect and analyze the 40 installed mobile applications. The results demonstrated that the applications had significant problems in regard to privacy policies, permission applications, information collection, and data storage. Likewise, Yu et al. \cite{yu2016can} also analyzed the privacy policies of Android applications by performing a systematic study that automatically identified three kinds of problems in privacy policies. They proposed a tool that successfully discovered that one in four applications has at least one type of policy-related issue. Papageorgiou et al. \cite{papageorgiou2018security} analyzed the security and privacy aspects of mobile health applications. The paper highlighted the concerning state of security practices in mobile health applications by manually investigating selected mobile health applications, which resulted in a finding that the majority of the analyzed applications did not follow the established practices, guidelines, and legal restrictions. Similarly, Ikram et al. \cite{ikram2016analysis} covered the domain of Android VPN (Virtual Private Network) permission-enabled applications. In their study, they analyzed the behavior of 283 applications regarding malware presence, third-party library embeddings, and traffic manipulation, including privacy aspects. The results showed that a significant number of applications use insecure VPN tunneling protocols.

\textit{Security awareness of mobile applications' users} - Zeybek et al. \cite{zeybek2019study} performed a user study wherein they investigated the security awareness of public institution personnel about the use of mobile devices. A similar study was conducted by Moletsane et al. \cite{moletsane2020mobile} that assessed factors that impact the mobile security awareness of students. The study proposed a conceptual model of mobile information security awareness that demonstrated a significant relationship between students' knowledge and behavioral intentions in case of threats from various security threats. Tao et al. \cite{tao2020identifying} proposed SRR-Miner, which is a review summarization approach that automatically retrieves security-related issues from user comments in mobile applications. This is achieved by applying methods such as a keyword-based search and deep analysis of sentence structures.

\begin{figure*}[h]
    \includegraphics[width=13cm]{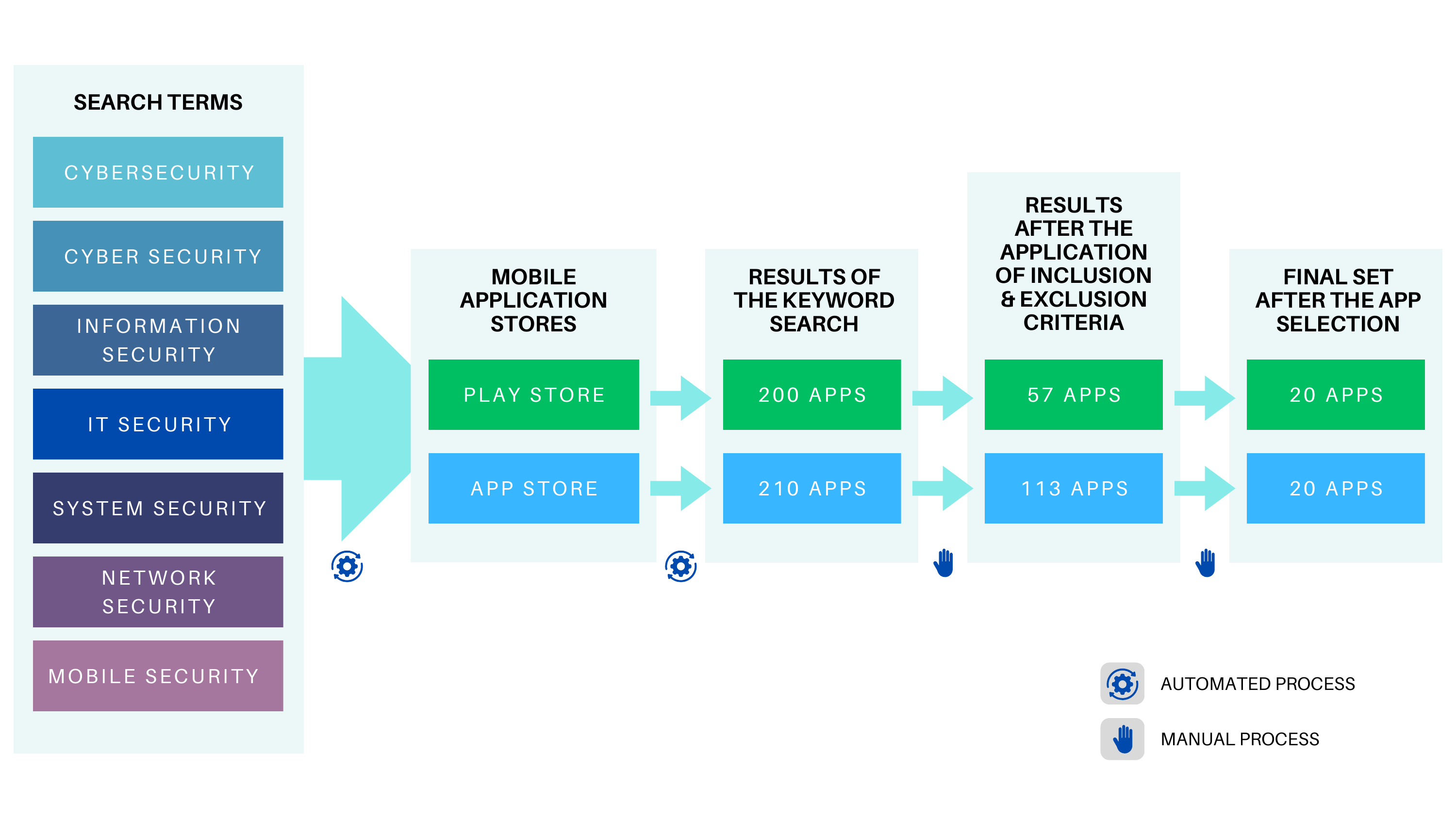}
    \caption{Overview of the Search and Selection Strategies}
    \label{fig:methodology}
\end{figure*}

The study that can be considered as the closest to this work is by Yao et al. \cite{yao2020security}. They conducted an empirical analysis of Android security applications by considering multiple aspects such as metadata, static analysis, and dynamic analysis. In addition, they also provided insights into the behaviors and operations of the analyzed applications. However, these aspects are completely different compared to what this study addresses\footnote{We also wanted to investigate if there were any common applications that were analyzed both by Yao et al. \cite{yao2020security} and this study. However, Yao et al. \cite{yao2020security} did not openly publish their artifacts.}. Our motivation behind this study was to gain a comprehensive overview of security applications and to explore their potential for cyber threat intelligence sharing. Thus, we focus on the type of security information the applications provide as well as detailed information on specific use cases of each application based on the NIST Cyber Security Framework \cite{barrett2018framework}. In addition, we also consider applications from the App Store as well as cover some other aspects that are discussed in detail in Section \ref{subsec:extraction}.  

\section{Proposed Method}
\label{sec:method}
\noindent

\noindent In this section, we present the design of our systematic review of mobile security applications (see Figure \ref{fig:methodology}). We outline our search strategy (see Section~\ref{subsec:search}), selection of mobile security applications (see Section~\ref{subsec:selection}), and their analysis (see Section~\ref{subsec:extraction}).

\subsection{Search Strategy}
\label{subsec:search}
\noindent To identify relevant mobile applications in the security domain, a keyword search on both App\footnote{\url{https://www.apple.com/at/app-store/}} and Play Store was conducted. We considered only these two as they are the two most popular mobile application stores. The search process is somewhat different compared to conventional systematic literature reviews \cite{kitchenham2004procedures} wherein a single string is formed by combining keywords with logical operators. In this study, multiple standalone search strings had to be formed because neither the App nor the Play Store supports advanced searches. Thus, the following seven search strings were formed: ``Cybersecurity'', ``Cyber Security'', ``Information Security'', ``IT Security'', ``System Security'', ``Network Security'', and ``Mobile Security''. We tested various keyword combinations and opted at the end for the aforementioned general strings because, in this case, the results of the searches included more relevant applications\footnote{If the search strings were too specific, many applications were not found, and in some cases, fewer than ten applications were identified.}. 

Given that multiple searches had to be conducted (i.e. for each specific string on both application stores) and that both stores advertise specific applications (i.e. which we wanted to avoid), we opted for an automated search by crafting a Python 3 script. The script utilizes \textit{google\_play\_scraper}\footnote{\url{https://pypi.org/project/google-play-scraper/}} library for the Play Store search and direct \textit{iTunes}\footnote{\url{https://itunes.apple.com/search}} search using the \textit{requests}\footnote{\url{https://pypi.org/project/requests/}} library for the App Store applications. We searched for the top 30 applications for each search string on each of the stores. This is because the Play Store search only provides 30 results and we wanted to keep the search strategy consistent. The results of the search were recorded in two separate spreadsheets wherein each sheet included seven tabs for each specific search term. Moreover, our scripts automatically provided meta information (e.g., \textit{cost}, \textit{number of downloads}, \textit{user rating}, etc.), which were utilized for selection of relevant applications.

\begin{table*}[!t]
\captionsetup{justification=centering}
\caption{Overview of Identified Mobile Security Applications: Functions, Use Cases, Provided Security Information, and Popularity based on Downloads or Ratings. (Note: I = Identify, P = Protect, D = Detect, R = Respond, N = None, R' = Recover, UC = Use Case, A = Asset, A'=Attack, C = Countermeasure, R = Risk, T = Threat, V = Vulnerability, O = Other, DL = \# of Downloads for GPL Applications, RT = \# of Ratings for IOS Applications). The Table is Sorted According to the Use Cases}
\begin{tabular}{@{\hspace{1cm}}llllllllllllllllll@{\hspace{1cm}}}
\toprule
                              &                                     & \multicolumn{7}{c}{\textbf{Functions addressed}}                                           &                               & \multicolumn{7}{c}{\textbf{Security Information}}                                         & \textbf{Popularity} \\
\multirow{-2}{*}{\textbf{ID}} & \multirow{-2}{*}{\textbf{App Name}} & \textbf{I} & \textbf{P} & \textbf{D} & \textbf{R} & \textbf{R'} & \textbf{G} & \textbf{N} & \multirow{-2}{*}{\textbf{UC}} & \textbf{A} & \textbf{A'} & \textbf{C} & \textbf{R} & \textbf{T} & \textbf{V} & \textbf{O}  & \textbf{\textcolor{table_1_blue}{DL}/\textcolor{table_1_green}{RT}}\\ \midrule
GPL01                         & Ethical Hacking University App      &            &            &            &            &             &            & x          & UC1                           & x          & x           & x          & x          & x          & x          &             & \textcolor{table_1_blue}{680k}\\
GPL03                         & Learn Ethical Hacking: HackerX      &            &            &            &            &             &            & x          & UC1                           &            & x           &            &            & x          & x          &             & \textcolor{table_1_blue}{6.5M}\\
GPL05                         & IT Cybersecurity Pocket Prep        &            &            &            &            &             &            & x          & UC1                           & x          & x           & x          & x          & x          & x          &             & \textcolor{table_1_blue}{162k}\\
GPL09                         & Learn Ethical Hacking               &            &            &            &            &             &            & x          & UC1                           & x          & x           & x          & x          & x          & x          &             & \textcolor{table_1_blue}{5.2M}\\
GPL18                         & Learn Cyber Security                &            &          &            &            &             &            &    x        & UC1                           & x          & x           & x          & x          & x          & x          &            & \textcolor{table_1_blue}{739k}\\
IOS01                         & CompTIA Security+ Exam Prep         &            &           &            &            &             &            & x           & UC1                           & x          & x           & x          & x          & x          & x          &            & \textcolor{table_1_green}{5.8k}\\
IOS03                         & CBT Nuggets                         &            &           &            &            &             &            &   x         & UC1                           & x          & x           & x          & x          & x          & x          &            & \textcolor{table_1_green}{5.8k}\\
IOS09                         & IT Cybersecurity Pocket Prep        &            &          &            &            &             &            &    x        & UC1                           & x          & x           & x          & x          & x          & x          &            & \textcolor{table_1_green}{4.8k}\\
GPL04                         & AVG AntiVirus Security              & x          & x          & x          & x          &             &            &            & UC2                           & x          & x           & x          & x          & x          & x          &            & \textcolor{table_1_blue}{465M}\\
GPL06                         & AVG Protection                      & x          & x          & x          &            &             &            &            & UC2                           & x          & x           & x          & x          & x          & x          &            & \textcolor{table_1_blue}{46M}\\
GPL07                         & ESET Mobile Security Antivirus      & x          & x          & x          & x          &             &            &            & UC2                           &            &             & x          & x          & x          &            &            & \textcolor{table_1_blue}{35M}\\
GPL08                         & VPN Antivirus by Kaspersky          & x          & x          & x          & x          &             &            &            & UC2                           & x          & x           & x          & x          & x          & x          &            & \textcolor{table_1_blue}{124M}\\
GPL10                         & Avast Antivirus Security            & x          & x          & x          &            &             &            &            & UC2                           & x          & x           & x          & x          & x          & x          &            & \textcolor{table_1_blue}{392M}\\
GPL11                         & Norton Genie: AI Scam Detector      & x          & x          & x          & x          &             &            &            & UC2                           & x          &             & x          & x          & x          &            &            & \textcolor{table_1_blue}{1.5M}\\
GPL12                         & Bitdefender Antivirus               & x          & x          & x          & x          &             &            &            & UC2                           & x          & x           & x          & x          & x          & x          &            & \textcolor{table_1_blue}{9.1M}\\
GPL13                         & Avira Security Antivirus VPN        & x          & x          & x          & x          &             &            &            & UC2                           & x          &             & x          & x          & x          &            &            & \textcolor{table_1_blue}{38M}\\
GPL14                         & Mobile Security Antivirus           & x          & x          & x          & x          & x           &            &            & UC2                           & x          & x           & x          & x          & x          & x          &            & \textcolor{table_1_blue}{6.2M}\\
GPL15                         & Bitdefender Mobile Security         & x          & x          & x          &            &             &            &            & UC2                           & x          &             & x          & x          & x          &            &            & \textcolor{table_1_blue}{17M}\\
GPL16                         & Norton360 Antivirus Security        & x          & x          & x          & x          &             &            &            & UC2                           & x          & x           & x          & x          & x          & x          &            & \textcolor{table_1_blue}{78M}\\
GPL20                         & Bitdefender Central                 & x          & x          & x          & x          & x           &            &            & UC2                           & x          & x           & x          & x          & x          & x          &            & \textcolor{table_1_blue}{1.6M}\\
IOS08                         & MSecure                             & x          & x          & x          &            &             &            &            & UC2                           &            &             & x          & x          &            &            & x          & \textcolor{table_1_green}{47.7k}\\
IOS16                         & Avast Security Privacy              &            & x          & x          &            &             &            &            & UC2                           & x          & x           & x          & x          & x          & x          &            & \textcolor{table_1_green}{24.6k}\\
IOS20                         & Norton360 Mobile Security, VPN      & x          & x          & x          & x          &             &            &            & UC2                           & x          & x           & x          & x          & x          & x          &            & \textcolor{table_1_green}{123.5k}\\
GPL02                         & Phone Guardian VPN: Safe WiFi       &            & x          & x          &            &             &            &            & UC3                           & x          &             &            &            &            &            &            & \textcolor{table_1_blue}{33M}\\
GPL19                         & Trustd Mobile Security              & x          & x          & x          &            &             &            &            & UC3                           & x          &             & x          & x          & x          &            &            & \textcolor{table_1_blue}{332k}\\
IOS04                         & Duck Duck Go                        &            &   x         & x           &            &             &            &           & UC3                           & x          & x           & x          & x          & x          & x          & x          & \textcolor{table_1_green}{4.1k}\\
IOS05                         & Google Authenticator                &            &    x        &            &            &    x         &            &           & UC3                           & x          & x           & x          & x          & x          & x          & x          & \textcolor{table_1_green}{12k}\\
IOS10                         & Surfshark VPN: Fast Reliable        &            & x          & x          & x          &             &            &            & UC3                           &            &             & x          & x          & x          &            &            & \textcolor{table_1_green}{86.6k}\\
IOS11                         & WireVPN - Fast VPN Proxy            &            & x          &            &            &             &            &            & UC3                           & x          &             & x          &            &            &            &            & \textcolor{table_1_green}{85.7k}\\
IOS12                         & Phone Guardian Safe Mobile VPN      &            & x          & x          &            &             &            &            & UC3                           &            & x           & x          &            & x          &            &            & \textcolor{table_1_green}{39.2k}\\
IOS17                         & Browsec VPN: Fast Ads Feed          &            & x          &            &            &             &            &            & UC3                           & x          & x           & x          & x          & x          &            &            & \textcolor{table_1_green}{16.4k}\\
IOS18                         & McAfee Security: Privacy VPN        &            & x          & x          &            &             &            &            & UC3                           & x          & x           & x          & x          & x          & x          &            & \textcolor{table_1_green}{175.8k}\\
IOS06                         & Safe Lock - Private Photo Vault     &            & x          & x          &            &             &            &            & UC4                           & x          &             & x          &            & x          &            &            & \textcolor{table_1_green}{23.4k}\\
IOS07                         & Private Photo Vault - Pic Safe      &            & x          & x          &            &             &            &            & UC4                           & x          & x           & x          & x          & x          &            &            & \textcolor{table_1_green}{1M}\\
IOS14                         & RoboForm Password Manager           &            & x          &            &            &             &            &            & UC4                           & x          &             & x          & x          &            &            &            & \textcolor{table_1_green}{43.2k}\\
IOS19                         & Hide Photos Video - Hide it Pro     &            & x          &            &            & x           &            &            & UC4                           & x          &             & x          & x          & x          &            &            & \textcolor{table_1_green}{46.1k}\\
GPL17                         & Fing - Network Tools                &            & x          & x          &            &             &            &            & UC5                           & x          &             &            & x          & x          & x          &            & \textcolor{table_1_blue}{62M}\\
IOS02                         & Duo Mobile                          & x          & x          & x          & x          &             & x          &            & UC5                           &            &             &            & x          & x          & x          &            & \textcolor{table_1_green}{4.8k}\\
IOS15                         & WebSSH - SysAdmin Tools             &            & x          &            &            &             &            &            & UC5                           & x          &             & x          &            &            &            &            & \textcolor{table_1_green}{0.6k}\\
IOS13                         & Signal - Private Messenger          &            & x          &            &            &             & x          &            & UC6                           & x          &             & x          &            &            &            &            & \textcolor{table_1_green}{853k}\\ \bottomrule
\end{tabular}
\label{tab:overview}
\end{table*}

\subsection{Application Selection}
\label{subsec:selection}
\noindent The search strategy resulted in a total of n=190 applications for the Play Store and n=210 applications for the App Store. In the next step, we manually checked all the identified applications. The goal was to obtain the final list of the top 20 most relevant security applications for each of the stores. Consequently, the following inclusion and exclusion criteria were applied. All the applications that were not freely available and did not provide or relate to any security topics were excluded from the final set. Additionally, we discarded all applications with fewer than 10,000 downloads and an application rating below 3 on the Play Store. For the App Store, we excluded applications with fewer than 1,000 user ratings and an application rating below 3. Moreover, we eliminated all duplicates, resulting in a total of 57 applications for the Play Store and 113 applications for the App Store. Finally, we ordered the list of applications based on the ratings and selected the top 20 applications for each store as a part of the final set that will be utilized for further analysis.

The aforementioned search strategy and application selection process resulted in a diverse corpus of applications. This encompassed a wide range of applications, from educational tools for security certifications to actual security implementation applications such as antivirus software or VPNs. We acknowledge that classifying all these applications under the umbrella term ``mobile security applications'' may not be ideal, as it groups together applications with vastly different purposes. For instance, applications designed to help users study for security certifications (e.g., \textit{CBT Nuggets} or \textit{CompTIA Security+ Exam prep}) serve a fundamentally different purpose compared to applications that actively secure devices or protect user privacy (e.g., antivirus applications, VPNs, or privacy-focused browsers such as \textit{DuckDuckGo}). However, the reason for doing this is due to the fact that we wanted to provide an overview of mobile security applications including their use cases, functions, and the type of security information they provide. In order to achieve this, it is necessary to cover all the different security-related domains, which aligns with the goal of this study.

Furthermore, we also note that the provided functions vary significantly. Educational applications may not deal with the identification of security risks such as the case with antivirus applications. Thus, the security measures implemented in \textit{Norton360 Antivirus Security} would be expected to be more extensive than those in an exam preparation application, given the difference in their nature.

\subsection{Application Analysis}
\label{subsec:extraction}
\noindent In order to analyze and classify all the identified applications, two authors were assigned to each specific application. During this process, the relevant metadata about the applications, as previously mentioned, was automatically retrieved. Any information that could not be automatically obtained had to be supplemented by manually analyzing the respective documentation in the respective stores. Below, we list all the steps and the type of information that was obtained as a part of the analysis. 

In the first step, the following information was automatically collected for each application: a \textit{unique identifier}, the application's \textit{name}, its \textit{current version}, the \textit{cost}, the total \textit{number of downloads}, the \textit{average user rating}, the list of \textit{permissions} requested upon installation\footnote{We utilized the respective Application Store to identify permissions listed for each of the applications.}, the year of the application's \textit{first release}, the year of its \textit{last update}, the \textit{types of notifications}, and whether a \textit{desktop version} is available.

In the second step, to address RQ1, we extracted and investigated detailed information on the specific \textit{use case} (UC) of each application and how it supports the \textit{functions} of the NIST Cyber Security Framework \cite{barrett2018framework}. In doing so, we differentiated among the following six functions:
\begin{itemize}[leftmargin=10pt]
    \item \textit{Identify} (I) - The application identifies current security risks and vulnerabilities in the system.
    \item \textit{Protect} (P) - The application offers protective measures and controls for protection against threats.
    \item \textit{Detect} (D) - The application provides information about threats detected on the system.
    \item \textit{Respond} (R) - The application provides suggestions on how to respond to a threat or improve the system's security.
    \item \textit{Recover} (R') - The application offers mechanisms to restore data and the system to a previous state. 
    \item \textit{Govern} (G) - The application provides information on security regulations and certifications.
    \item \textit{Other} (O) - The application does not support any of the functions (i.e. I, P, D, R, R', or G) mentioned in the NIST CSF.
\end{itemize}

In a third step, to address RQ2, we extracted and examined the \textit{type of information} the application provides on various security aspects. In doing so, we differentiated among the following six information types \cite{iso2009iso}:
\begin{itemize}[leftmargin=10pt]
    \item \textit{Asset} (A) - Information about sensitive information or services of the system.
    \item \textit{Attack} (A') - Information about a deliberate form of compromise, i.e. an unwanted or unauthorized act to cause damage to assets or systems.
    \item \textit{Countermeasure} (C) - Information about a measure or technique that reduces the impact of a threat or vulnerability by eliminating or preventing it.
    \item \textit{Risk} (R) - Information about a potential threat or vulnerability that may compromise the confidentiality, integrity, or availability of the assets or the system.
    \item \textit{Threat} (T) - Information about the potential cause of an attack that could damage the assets or the system.
    \item \textit{Vulnerability} (V) - Information about an asset's or system's weaknesses that can be exploited by a threat.
    \item \textit{Other} (O) - The application provides other types of information than those (i.e. A, A', C, R, T, V, and O) mentioned.
\end{itemize}

After completing all the classification tasks, we measured the inter-coder reliability score. This was done for all the 40 applications. We observed an agreeability score of 94\%, denoting that the authors likely reached the same conclusion. Finally, any classification discrepancies were resolved through in-depth discussions. The aforementioned procedure resulted in a spreadsheet containing classified applications with all the relevant information, which we will discuss in detail in Sections~\ref{sec:results} and \ref{sec:discussion}.

\section{Results}
\label{sec:results}
\noindent

\noindent In this section, we present the results of our systematic study of mobile security applications (see Table \ref{tab:overview}). We provide a general overview (see Section~\ref{subsec:general}), discuss their use cases (see Section~\ref{subsec:usecase}) and the security information (see Section~\ref{subsec:security_information}) they provide.

\subsection{Overview of Mobile Security Applications}
\label{subsec:general}
\noindent \textit{Overview.} Table \ref{tab:overview_apps} provides the statistical overview of the 40 analyzed applications\footnote{We could not directly compare the application usage numbers between the two stores because the download rates for App Store applications were unavailable. Therefore, we used the rating count as a proxy.}. Our investigation revealed that 24 applications have a dedicated desktop version, and 7 applications offer a web version. Only about 20\% of the applications are exclusively available on mobile devices. The average rating of all the analyzed applications is 4.72 out of 5. This indicates that we only focused on the best-rated applications. Furthermore, we reviewed which store tag each app was assigned to. The results indicated a tendency toward the \textit{Tools}, \textit{Education}, and \textit{Utilities} categories. It is important to note that the Google Play Store and the App Store do not always use identical tags. Google Play uses the tag \textit{Tools}, whereas the App Store refers to this category as \textit{Utilities}.



\begin{table}[h]
\centering
\caption{Overview of Applications' Metrics and Statistics}
\resizebox{\columnwidth}{!}{
\begin{tabular}{>{\raggedright\arraybackslash}p{4.5cm}>{\raggedright\arraybackslash}p{5.5cm}}
\toprule
\textbf{Description} & \textbf{Value} \\
\midrule
Number of apps per store & 20 GPL, 20 IOS \\ 
Average \# of downloads GPL & 63,574,897.45 \\
Total \# of downloads GPL & 1,271,497,949 \\
Average ratings count for IOS & 290,303.05 \\
Total ratings count for IOS & 5,806,061 \\
Average rating (both stores) & 4.7232 stars \\
Other app implementations & Desktop: 24, Web: 7, Mobile-only: 9 \\
Maintenance & Actively maintained: 37, Inactive: 3 \\
Apps have a paid option & Yes: 34, No: 6 \\
Provided information counts & Countermeasures: 36, Assets: 34, Threats: 34, Risks: 33,  Attacks: 24, Vulnerabilities: 23, Others: 3 \\
CSF: Use-Case counts & Protect: 32, Detect: 25, Identify: 16, Respond: 12, Recover: 4, Govern: 2, None: 8 \\
\bottomrule
\end{tabular}
}
\label{tab:overview_apps}
\end{table}

\noindent \textit{Timeline.} Another aspect that we investigated was the timeline of applications starting from their first release to the last update. According to Figure \ref{fig:app_activity_maintained}, the large majority of all applications are regularly updated and maintained (n=36), while 4 applications did not receive any updates in the last six months and these can be considered inactively managed, which can potentially create security-related issues for the users of these applications.


\begin{figure*}[h]
\includegraphics[width=0.855\textwidth]{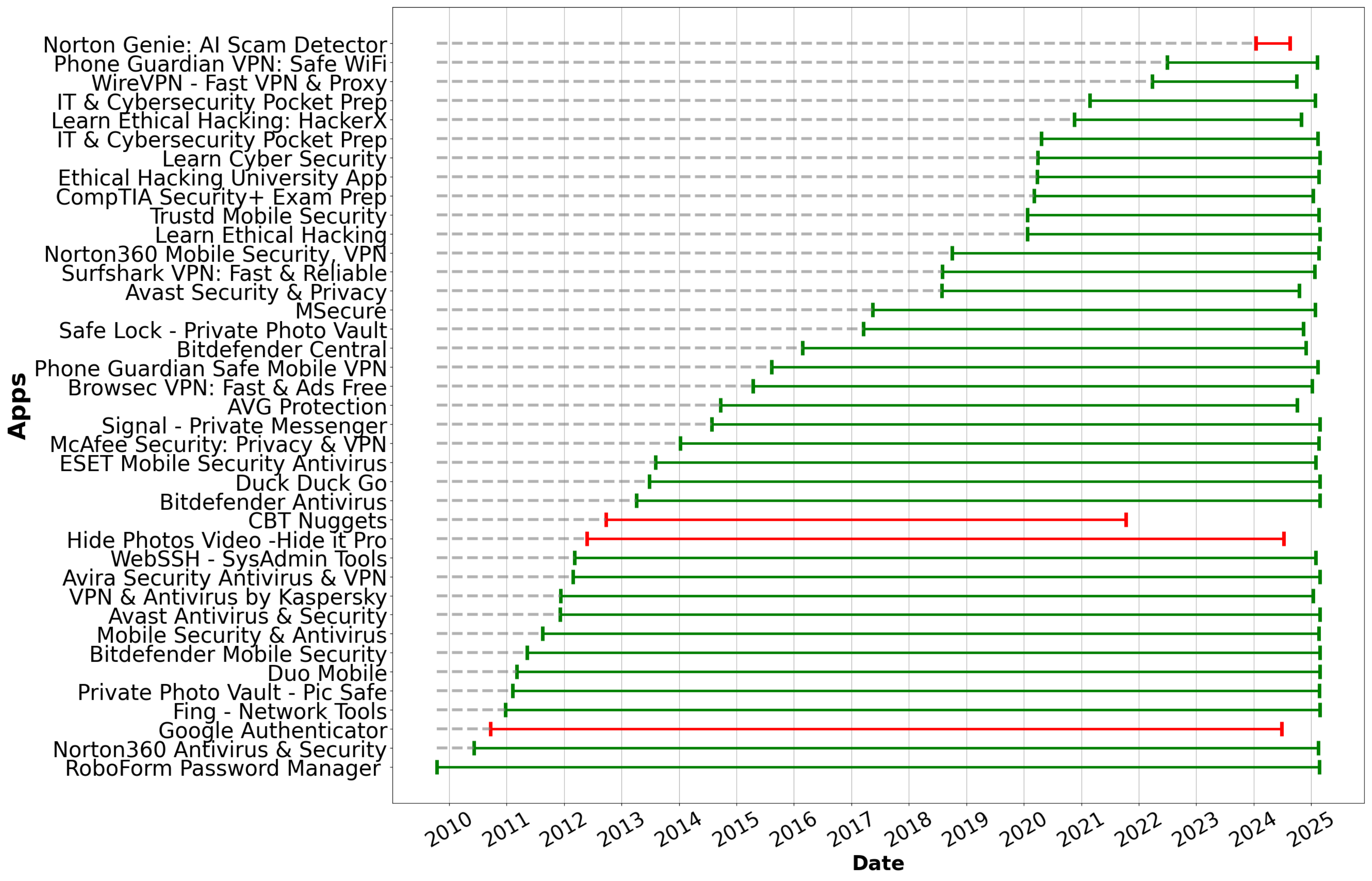}
    \caption{Timeline of Applications' Release Until Last Updates}
\label{fig:app_activity_maintained}
\end{figure*}



\noindent \textit{Popularity.} In order to assess the popularity of the analyzed applications, we compared the number of users (downloads for GPL and rating counts for IOS) with their average ratings\footnote{For readability purposes, the number of downloads and the number of ratings is represented as $n \times 10^8$ and  $n \times 10^6$ respectively.}. Figure \ref{fig:popularity_downloads_rating} presents this comparison, with the GPL apps displayed on the left y-axis and the iOS applications on the right y-axis. In addition, the regression lines for both platforms, along with confidence intervals are demonstrated using red and blue shading. The plot shows a positive correlation between a higher number of users and better ratings, which is more evident for the IOS Applications compared to the GPL applications.

\begin{figure}[h]
\includegraphics[width=0.9\columnwidth]{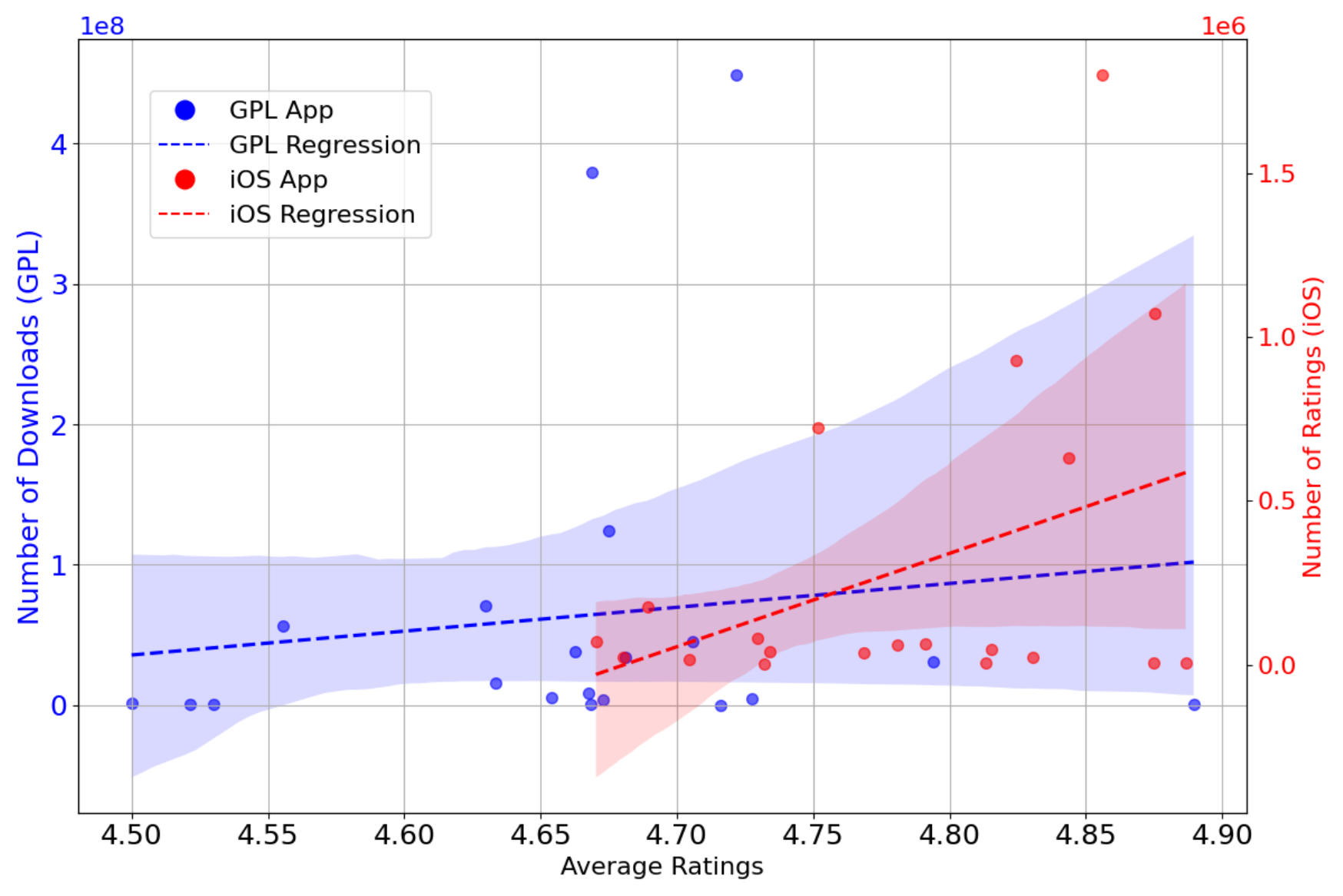}
    \caption{Number of Users vs. Average Ratings}
\label{fig:popularity_downloads_rating}
\end{figure}


\noindent \textit{Permissions.} Another general aspect that we considered was the applications' permissions. Figure \ref{fig:permissions_barchart} presents the distribution of permissions listed on the store page, required to utilize the full functionality of the examined applications. The number of required permissions averaged 3.575 per application. For simplicity, the analysis was limited to these basic listed permission categories, excluding specific and special permissions. This is due to the fact that specific and special permissions can indeed be too particular and differ for every application, making it very difficult to list all of these. For example, this can be permissions needed to schedule exact alarms or permissions to display and draw over other applications\footnote{\url{https://developer.android.com/training/permissions/requesting-special?hl=en}}.  Our results demonstrated that the most demanded permissions include \textit{Location} (50\%) and \textit{Storage} (47.5\%) permissions. This is followed by \textit{Photos/Media/Files} (35\%), \textit{Contacts} (35\%), \textit{Identity} (27.5\%), and \textit{Camera} (27.5\%) permissions. We only present the permissions that were identified four or more times. The detailed list of all permissions needed by each application we analyzed can be found in our publicly shared repository.

\begin{figure}[h]
    \includegraphics[width=\columnwidth]{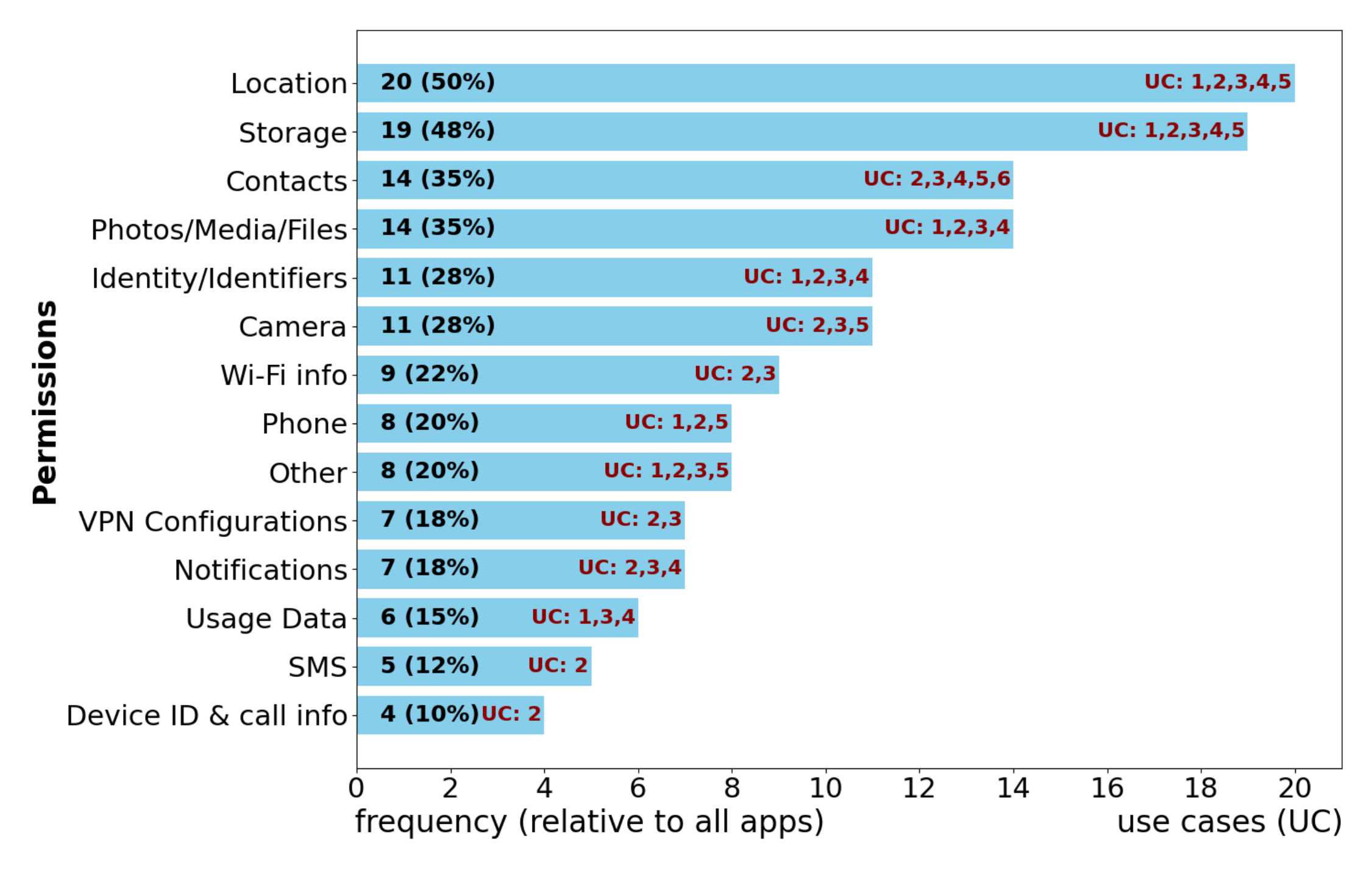}
    \caption{Distribution of Required Permissions and Use-cases}
\label{fig:permissions_barchart}
\end{figure}

\subsection{Use Cases of Mobile Security Applications (RQ1)}
\label{subsec:usecase}
\noindent As depicted in Table~\ref{tab:overview}, our investigations showed that the core use cases of mobile security applications are \textit{Security Education \& Training} (UC1), \textit{Antivirus Protection} (UC2), \textit{Secure Browsing} (UC3), \textit{Privacy Management} (UC4), \textit{Network \& Device Management} (UC5) and \textit{Secure Communication} (UC6). In the following, we discuss these use cases by providing examples of selected applications\footnote{Due to space constraints, we could not include all identified applications as examples in our explanations.}:\\

\textit{Security Education \& Training} (UC1) applications are designed to teach users the basics and advanced concepts of security and ethical hacking and prepare them for professional certifications. For example, the \textit{Ethical Hacking University} application, offers interactive courses that teach users how to protect digital assets by recognizing, preventing, and responding to cyber threats. Similarly, the \textit{Learn Ethical Hacking: HackerX} application offers hands-on exercises and interactive modules to teach ethical hacking techniques and security skills. For those seeking industry-recognized certification, the \textit{IT \& Cybersecurity Pocket Prep} application provides comprehensive practice questions and learning materials. In addition, the \textit{CBT Nuggets} application offers an extensive library of training courses accessible on mobile devices including various videos taught by experts, which can be downloaded for offline learning. 

\textit{Antivirus Protection} (UC2) applications focus on protecting mobile devices from a wide range of threats, including malware, phishing, and unauthorized access. For example, the \textit{AVG AntiVirus} application is a comprehensive security solution that includes real-time virus and malware scanning, application blocking, encrypted photo storage, VPN protection, Wi-Fi threat detection, and hacker alerts. Another example is the \textit{ESET Mobile Security Antivirus} application, which protects against malware, phishing attempts, and other digital threats to ensure a safe mobile experience. The \textit{Bitdefender Mobile Security} application offers comprehensive protection against malware, phishing attacks, and privacy threats. Similarly, the \textit{Norton 360 Mobile Security \& VPN} application combines malware and phishing protection with secure VPN services to improve both security and privacy for mobile users. 

\textit{Secure Browsing} (UC3) applications are designed to enable secure internet connections and protect users' online privacy. For example, the \textit{Surfshark VPN} application encrypts online traffic and masks IP addresses so that users can surf the Internet safely. The \textit{Phone Guardian Mobile Security} application provides a secure VPN connection and alerts users to potential security threats to keep their data private. The \textit{Browsec} application enables encrypted web browsing and access to multiple virtual locations, improving user privacy. The \textit{DuckDuckGo} application focuses on secure and private online searches by blocking trackers and ensuring that users' search histories and personal data remain confidential. 

\textit{Privacy Management} (UC4) applications help users to securely store and manage their passwords and protect their private photos and videos. For example, the \textit{mSecure} application enables users to manage and store their passwords and personal data with by utilizing efficient cryptographic algorithms such as AES-256 Encryption to protect their privacy. The \textit{RoboForm Password Manager} application securely stores and manages all passwords and login credentials and enables quick and secure access to online accounts across multiple devices. To protect sensitive photos and videos, the \textit{Safe Lock - Private Photo Vault} application offers a wide range of security features such as PIN codes, Touch ID, and intrusion alerts. With the \textit{Hide it Pro} application, users can securely hide their photos and videos behind a screen lock, with additional privacy features such as customizable albums and escape unlock codes. 

\textit{Network \& Device Management} (UC5) applications provide tools to efficiently manage, monitor, and secure networks and devices to ensure optimal performance and security. For example, the \textit{Fing - Network Tools} application helps users manage and secure their home and business networks by providing detailed insights into connected devices, network performance, and security vulnerabilities. The \textit{WebSSH Essential} application offers robust features for secure remote server management, including SSH, SFTP, TELNET, port forwarding, and secure access via Touch ID or Face ID. The \textit{Duo Mobile} application enhances enterprise security through multi-factor authentication, user identity verification, and protection against unauthorized access to applications and systems. 

\textit{Secure Communication} (UC6) applications prioritize user privacy by offering encrypted messages and calls. For example, the \textit{Signal} application offers end-to-end encrypted texts, voice \& video calls, and group chats while ensuring that any data is collected, maintaining user privacy and security. This application is particularly popular with users who are conscious of their digital privacy and want to ensure that communications remain confidential and protected from unauthorized access. 


\begin{figure}[h]    \centering\includegraphics[width=0.9\columnwidth]{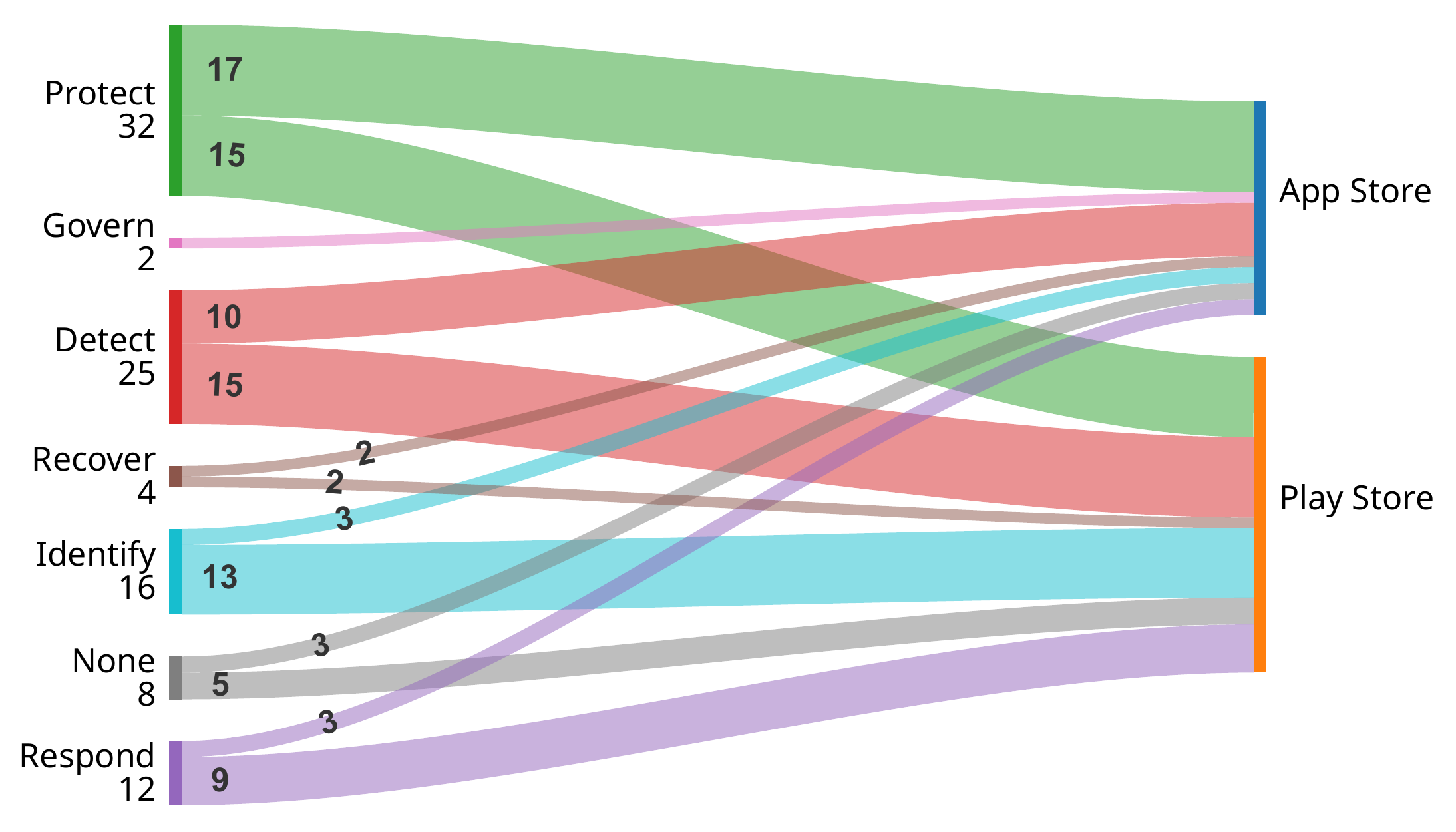}
    \caption{Overview of the Key Security Functions for Each of the Stores (for all the Selected Mobile Applications)}
    \label{fig:sankey1}
\end{figure}

Figure~\ref{fig:sankey1} (left side) shows the analysis on how various mobile applications support key security functions. We found that 32 applications offer protective measures and controls to guard against threats encompassing both organizational and technical aspects(i.e. \textit{Protect)}. Another 25 applications include functions for monitoring system threats (i.e. \textit{Detect}). 16 applications focus on identifying security risks and system vulnerabilities (i.e. \textit{Identify}). Additionally, a small set provides functionalities for incident response (i.e. \textit{Respond}; 12 applications) and recovery (i.e. \textit{Recover}; 4 applications). Finally, only 2 applications support governance processes (i.e. \textit{Govern}).

As shown in Figure~\ref{fig:sankey1} (right side), a comparison of the applications available in the App Store and Play Store reveals that functions for \textit{Identify} are scarcely offered in App Store applications. In contrast, the Play Store exhibits a balanced distribution among the \textit{Identify}, \textit{Protect}, and \textit{Detect} functions. The \textit{Respond}, \textit{Recover}, and \textit{Govern} functions are similarly under-supported in applications from both stores.

Subsequently, we also investigated the relationship between specific use cases and requested permissions (see Figure \ref{fig:permissions_barchart}). The results show the following associations: UC1 $\rightleftarrows$ {} location, storage, photos/media/files, identity, phone, usage data permissions; UC2 $\rightleftarrows$ {} location, storage, contacts, photos/media/files, identity, camera, WIFI information, phone, VPN configurations, notifications, SMS, device ID and call information permissions; UC3 $\rightleftarrows$ {} location, storage, contacts, photos/media/files, identity, camera, WIFI information, VPN configurations, notifications permissions; UC4 $\rightleftarrows$ {} location, storage, contacts, photos/media/files, identity, notifications, usage data permissions; UC5 $\rightleftarrows$ {} location, storage, contacts, camera, phone permissions; UC6 $\rightleftarrows$ {} contacts permissions. 


\subsection{Provided Security Information (RQ2)}
\label{subsec:security_information}

\noindent A comprehensive analysis of various types of security information provided by mobile applications revealed several interesting findings (see Figure \ref{fig:typesec}). Notably, \textit{vulnerability} and \textit{attack} information were found to be the least prevalent in these applications, primarily being provided by antivirus applications such as \textit{AVG Protection} or \textit{Avast Security Privacy}. Another observation indicated that applications from the Play Store exhibit a more security-oriented approach. However, when it comes to \textit{countermeasures}, applications from the App Store (e.g. \textit{Phone Guardian Safe Mobile VPN}, \textit{Duck Duck Go}, etc.) are more represented compared to Play Store applications. This discrepancy could be attributed to the higher identification of antivirus applications in the Play Store, thus affecting the distribution of security-oriented content. The most striking contrast between the Play Store and the App Store lies in the representation of \textit{threat}-related information such as \textit{Bitdefender Antivirus} or \textit{Trustd Mobile Security}. Notably, \textit{threat} information was found to be present in the Play Store 95\% of the time, whereas it was identified in 75\% of the applications in the App Store.

In general, the distribution of other types of security information appeared to be relatively uniform between the Play Store and the App Store, indicating a balanced approach to disseminating security-related content. Furthermore, the analysis indicated that 60\% of Play Store applications and 35\% of App Store applications provide all types of security information. In fact, these apps were predominantly categorized as educational applications (e.g. \textit{Ethical Hacking University} application, \textit{CompTIA Security+ Exam Prep}, etc.) and antivirus applications, wherein their role is to offer comprehensive security guidance and protection. Finally, the notable classification of three applications (\textit{Duck Duck Go}, \textit{Google Authenticator}, \textit{MSecure}) in the App Store as \textit{others} was due to their provision of general security advice, such as data backup and software update recommendations.

\begin{figure}[h]
\pdfinclusioncopyfonts=1
\includegraphics[width=0.9\columnwidth]{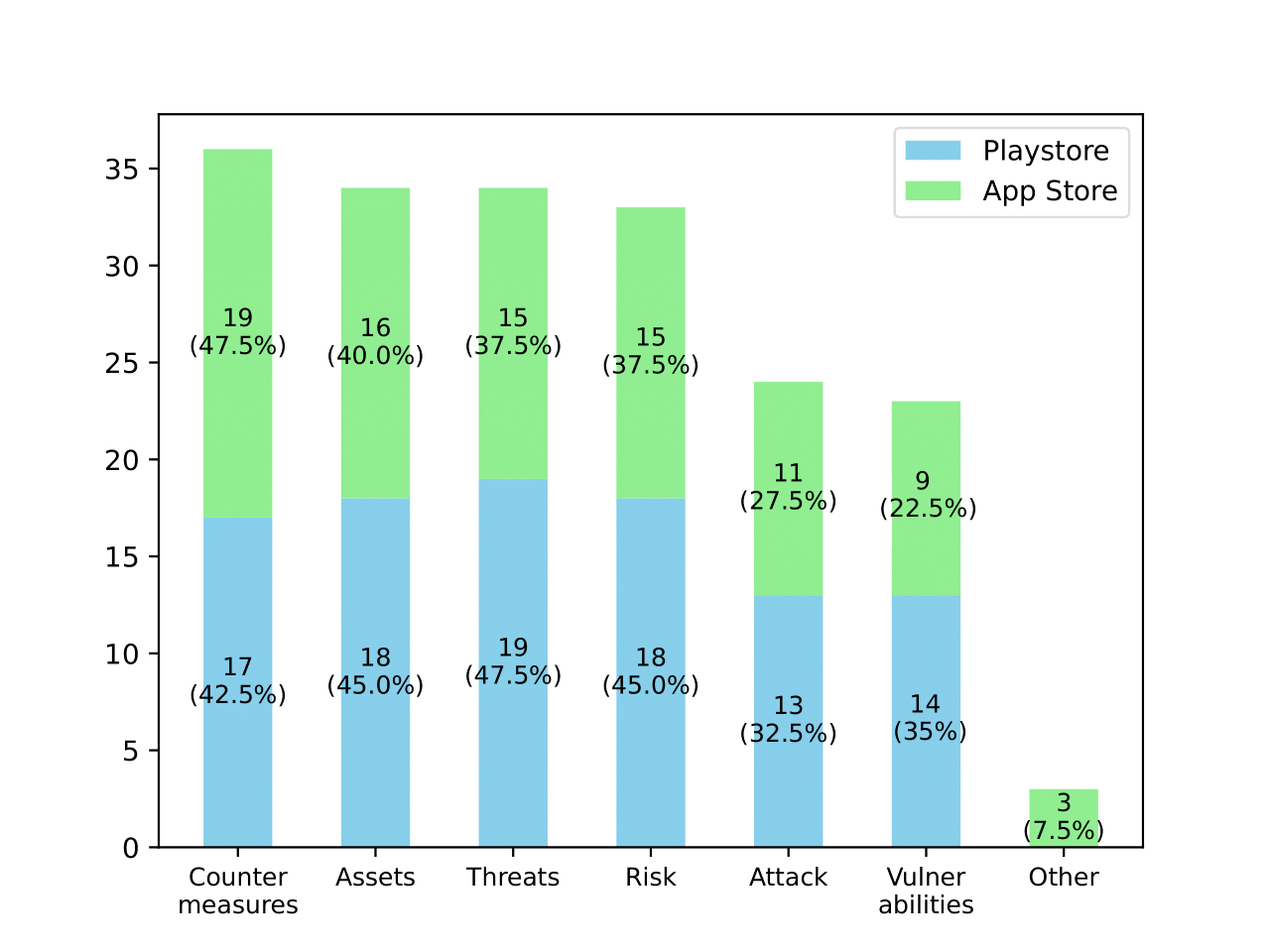}
    \caption{Type of Security Information for Each Store}
\label{fig:typesec}
\end{figure}
\section{Discussion}
\label{sec:discussion}
\noindent

\noindent In this section, we discuss key findings (see Section~\ref{subsec:keyfindings}) related to our research questions (RQ1 and RQ2), application development recommendations (see Section~\ref{subsec:recommendations}) and the limitations of our systematic analysis of mobile security applications (see Section~\ref{subsec:limits}).

\subsection{Key Findings and Implications}
\label{subsec:keyfindings}
\noindent Our research indicates that a significant number of mobile security applications are highly popular among users. Notably, applications with a large user base tend to have significantly better ratings, a trend that is particularly pronounced for iOS applications. Additionally, we discovered that many of these mobile security applications also offer desktop versions. This is often because they are extensions of established antivirus software (UC2). Furthermore, most of these applications provide paid features, reflecting the substantial business opportunities they represent.

A more detailed analysis of application store tags and our identified use cases reveals that a significant proportion of popular mobile applications incorporate essential functions (cf. store tag "utilities") for threat detection and protection \cite{damaraju2021mobile}. This trend is partly due to the prevalence of antivirus protection applications (UC2). Additionally, our research highlights the widespread availability of mobile educational tools (cf. store tag "education") aimed at enhancing security awareness and training (UC1)\footnote{This corresponds to the findings by Drigas and Angelidakis \cite{drigas2017mobile} that mobile applications have a potential to take education out of classroom boundaries.}. This underlines the importance of ongoing efforts to inform users about potential risks and provide practical guidance on mitigating various threats. 

Furthermore, we have identified numerous tools (cf. store tag "tools") offering features such as secure browsing (UC3), privacy management (UC4), network \& device management (UC5), and secure communication (UC6). Given that iOS and Android platforms also provide some of these security functions, it raises the question of why users opt for additional applications. This question is particularly relevant as our research indicates that these applications often require a significant number of permissions (as indicated by Gruschka et al. \cite{gruschka2018classification}), despite their focus on security and privacy, which necessitates a high level of user trust. Accordingly, a further question arises as to whether users place more trust in platform manufacturers or dedicated application providers in terms of security and privacy --- especially because some studies showed that there is a misuse of private data that third-party applications are collecting \cite{nguyen2022freely}. While our study hints at the importance of user trust in this decision-making process, a more thorough investigation is needed to fully understand why users might choose third-party security applications over native security features. This could involve exploring various aspects and factors such as perceived effectiveness, additional features, brand reputation, or even misconceptions about built-in security measures. 

On top of that, there is a significant relationship between permissions requested by mobile applications and user trust as well as popularity. Notably, applications that requested fewer permissions were often rated better by users. This suggests that privacy concerns play a critical role in application selection. This highlights an important need for developers in the community to minimize unnecessary permissions and communicate to users why specific permissions are required. This can be achieved by adopting a more rigorous permissions model, which should enhance user trust and encourage compliance with privacy best practices. Similarly, the features provided by popular applications can directly impact user engagement and perceived effectiveness. Our findings indicate that applications offering functionalities such as Privacy Management and Secure Communication tend to attract more users, which demonstrates the importance of integrating users' privacy needs into the design process. This correlation points out a crucial area for future development within the usable security field, where features should prioritize more user privacy. 

Moreover, a closer analysis of the security information provided by these applications revealed that contrary to our initial assumptions, information about countermeasures, assets, threats, and risks is more common than details on specific attacks and vulnerabilities. This can be considered unusual because vulnerability information is crucial and utilized in a wide range of different cybersecurity approaches \cite{groner2023study}. However, this finding highlights a significant opportunity for improvement in mobile security application development. Developers need to consider incorporating features that educate users about existing vulnerabilities in their devices and applications, as well as how to address them. Empowering users to take proactive measures to secure their devices is essential for enhancing overall mobile security.

Lastly, the majority of applications offer a variety of security information types. Surprisingly, we did not identify any cyber threat intelligence sharing mobile applications, which is a very unexpected finding given the current cybersecurity landscape --- including the rise of cyber threat intelligence sharing platforms \cite{sun2023cyber}. The absence of sharing capabilities is particularly concerning given the increasing complexity of cyber threats, which often require collaboration and community-driven approaches. This gap suggests the need to create such platforms that enable users to share insights and experiences related to mobile and any other types of security threats. This capability could become increasingly important in the future, as mobile security applications have the potential to act as security sensors, thereby enabling a form of threat intelligence sharing.

\subsection{Development Recommendations}
\label{subsec:recommendations}
\noindent
Our findings emphasize the need for mobile security application developers to define \textbf{clear use cases} that align with user needs across critical areas such as Security Education \& Training, Antivirus Protection, and Secure Communication. Providing \textbf{comprehensive security information} is significant so users can recognize potential threats and make informed decisions regarding their security posture. A focus on \textbf{user-centric design} should enhance user experience, while \textbf{regular updates} help to adapt to emerging vulnerabilities. Furthermore, integrating mobile security applications with \textbf{native features of operating systems} can lead to a more unified and consistent user experience.

Additionally, promoting a culture of \textbf{cyber threat intelligence sharing} can improve general security awareness among users. Adopting a \textbf{privacy-first approach}, which involves transparent data collection practices and encryption, is critical for user trust. This means that engaging users actively and seeking their feedback will lead to \textbf{continuous improvement}, ensuring that applications evolve with respect to real-world challenges. Overall, these recommendations should improve user engagement and contribute to a more secure mobile ecosystem.



\subsection{Limitations}
\label{subsec:limits}
\noindent The research at hand might be limited by a (i) \textit{selection bias} of mobile security applications, (ii) \textit{incorrect application analysis}, and (iii) \textit{missing out relevant applications}. As described in Section~\ref{subsec:selection}, our selection strategy was systematic and based on certain criteria (e.g. number of downloads, license model, etc.) and the popularity of these applications as rated by users to ensure that (i) is minimized. Accordingly, the selection of applications was not based on individual decisions by the authors, but by the users of the respective applications. To prevent (ii), we opted for a type of cross-validation in which each author of this research paper had to analyze a subset of applications that overlapped with another author's set. In this way, classification discrepancies were detected early and limited by reclassification. Finally, there is a chance that we missed some relevant applications (iii). To address the retrieval limitations of the Play Store, which restricted our sample to 30 applications, we implemented a strategic approach. This involved us originally testing and afterward applying multiple search strings to ensure that the applications from various cybersecurity domains were identified.  

\section{Conclusions}
\label{sec:conclusions}

\noindent As a part of this study, we investigated the landscape of mobile security applications for Android and iOS platforms. This is achieved by systematically providing a comprehensive overview of the top 20 most used applications in each of the stores. As a part of the analysis, we employed both automated and manual techniques. The results underscore the critical need for ongoing research and development in this domain. By identifying common use cases and the prevalent types of security information, our findings offer valuable insights into security-related aspects of these applications. Presumably, the fact that no threat intelligence-sharing mobile applications were identified can be considered rather alarming. 

Consequently, as the mobile security application ecosystem continues to evolve, our study resources have been made openly accessible, inviting further exploration and contribution to this vital area of study. As a part of future work, we plan to investigate the reasons for using and selecting these applications regardless if some mobile devices provide the same built-in security features.
The question arises of whether users place more trust in platform manufacturers or dedicated application providers in terms of security and privacy. For example, this could be investigated by performing a user study on a targeted group of the population as shown by Braun et al. \cite{braun2024understanding}, which can include both questionnaires and interviews. In addition, we plan to extend this study by conducting the analysis on a larger number of applications and comparing their features to their dedicated desktop and web versions. 

Furthermore, another aspect for future research is to evaluate whether these mobile security applications enhance measures provided by mobile operating systems and whether the security gains outweigh the possible risks of third-party apps. This investigation would provide insights into the effectiveness and necessity of these applications in the broader context of mobile device security.

\section*{Acknowledgement}

\noindent This research was partially supported by Hilti.

\balance
\bibliographystyle{ACM-Reference-Format}

{
\bibliography{bibliography} 
}


\end{document}